\documentclass[aip, cha, reprint]{revtex4-1}
\usepackage{xspace,units}
\usepackage{graphicx, multirow}
\usepackage{amssymb, amsmath, amsfonts}
\usepackage{xcolor}
\usepackage{enumitem}

\newcommand{\iu}{{i\mkern1mu}}

\newcommand{\defi}{\mathrel{\mathop:}=}

\newcommand {\CE}{$\mathcal{C}^{\rm E}$\xspace}
\newcommand {\CC}{$\mathcal{C}^{\rm C}$\xspace}
\newcommand {\CB}{$\mathcal{C}^{\rm B}$\xspace}
\newcommand {\CEV}{$\mathcal{C}^{\rm E}_{\rm v}$\xspace}
\newcommand {\CEE}{$\mathcal{C}^{\rm E}_{\rm e}$\xspace}

\newcommand {\CCV}{$\mathcal{C}^{\rm C}_{\rm v}$\xspace}
\newcommand {\CCE}{$\mathcal{C}^{\rm C}_{\rm e}$\xspace}

\newcommand {\CBV}{$\mathcal{C}^{\rm B}_{\rm v}$\xspace}
\newcommand {\CBE}{$\mathcal{C}^{\rm B}_{\rm e}$\xspace}

\newcommand {\PER}{\ensuremath{\cal{P}}\xspace}
\newcommand {\MEDR}{\ensuremath{{R_{\rm m}}}\xspace}

\newcommand {\vplus}{\ensuremath{\text{v}_>}\xspace}
\newcommand {\vminus}{\ensuremath{\text{v}_<}\xspace}

\newcommand {\rmstrength}{\ensuremath{\tilde{s}}\xspace}
\newcommand {\NWebs}{\ensuremath{N_W}\xspace}

\begin{document}

\title{Centrality-based identification of important edges in complex networks}

\author{Timo Br\"ohl}
\email{timo.broehl@uni-bonn.de}
\affiliation{Department of Epileptology, University of Bonn Medical Centre, Sigmund-Freud-Stra\ss{}e 25, 53105 Bonn, Germany}
\affiliation{Helmholtz Institute for Radiation and Nuclear Physics, University of Bonn, Nussallee 14--16, 53115 Bonn, Germany}

\author{Klaus Lehnertz}
\email{klaus.lehnertz@ukbonn.de}
\affiliation{Department of Epileptology, University of Bonn Medical Centre, Sigmund-Freud-Stra\ss{}e 25, 53105 Bonn, Germany}
\affiliation{Helmholtz Institute for Radiation and Nuclear Physics, University of Bonn, Nussallee 14--16, 53115 Bonn, Germany}
\affiliation{Interdisciplinary Center for Complex Systems, University of Bonn, Br{\"u}hler Stra\ss{}e 7, 53175 Bonn, Germany}

\begin{abstract}
Centrality is one of the most fundamental metrics in network science.
Despite an abundance of methods for measuring centrality of individual vertices, there are by now only a few metrics to measure centrality of individual edges.
We modify various, widely used centrality concepts for vertices to those for edges, in order to find which edges in a network are important between other pairs of vertices. 
Focusing on the importance of edges, we propose an edge-centrality-based network decomposition technique to identify a hierarchy of sets of edges, where each set is associated with a different level of importance.
We evaluate the efficiency of our methods using various paradigmatic network models and apply the novel concepts to identify important edges and important sets of edges in a commonly used benchmark model in social network analysis as well as in evolving epileptic brain networks.  
\end{abstract}
\maketitle

\frenchspacing

\begin{quotation}
In a large number of natural and man-made networks, an edge represents some form of interaction between pairs of vertices. 
Examples include the climate, brain networks, protein-protein interactions, gene interactions, plant-pollinator interactions, food-webs, or communication and social networks, to name just a few.
Improving understanding of structure and function of such interaction networks requires knowledge about their key constituents. 
Important vertices (or groups thereof) can be identified with a multitude of vertex centralities, however, there are only a few centrality-based concepts and metrics to identify important edges (or groups thereof). 
Here we propose straightforward modifications of widely used centrality concepts for vertices to those for edges and present a modification of the concept of network decomposition that allows one to identify sets of interconnected important edges. 
Using various paradigmatic network models and real data sets, we show how key network constituents --~as identified with the same centrality concept~-- are related to each other and demonstrate how important edges highlight typical properties of network topologies.
\end{quotation}

\section{Introduction}
The last decade witnessed an exceptional success of network theory and its applications in diverse areas of science including physics, earth and climate sciences, sociology, quantitative finance, biology, and the neurosciences~\cite{boccaletti2006a,arenas2008,bullmore2009,donges2009b,allen2011,barthelemy2011,barabasi2011,newman2012,baronchelli2013,lehnertz2014,heckmann2015,gao2016}. 
A growing number of studies indicate that research into the dynamics of spatially extended complex systems benefits from a network approach, as it allows a refined exploration of relationships between structure and dynamics in such systems. 
With this approach, a system is described by an interaction network, whose vertices represent elementary units of the system and whose edges represent interactions between them. 
There is now increasing evidence that understanding of structure--dynamics relationships can further be improved by allowing vertices and/or edges to change with time, which leads to so-called evolving interaction networks~\cite{holme2012,belykh2014,kivela2014,lehnertz2014}.

Previous studies mostly investigated macroscopic network properties (such as clustering, efficiency, or assortativity) and reported non-trivial characteristics for various empirically derived interaction networks.
More recently, special emphasis has been placed on the mesoscopic structural organization (e.g., on communities~\cite{fortunato2016}, on motifs~\cite{alon2007,fornito2015}, or on core--periphery structure~\cite{rombach2014}) as well as on microscopic aspects, such as the role of individual network constituents (i.e., vertices or edges) in structure and dynamics of interaction networks~\cite{lue2016}.
It is clear that identifying key network constituents and characterizing their importance for structure and dynamics of interaction networks is highly relevant to improve understanding and controlling of the collective dynamics~\cite{havlin2014,bialonski2015,liu2016,gates2016,lehnertz2016}.

There are different concepts and a growing number of metrics --~such as centralities~-- that allow one to characterize the role of network vertices for structure and dynamics~\cite{kuhnert2012,spitz2014,almgren2016,lue2016,geier2017b}. 
Interestingly, there are by now only a few metrics that characterize the role of individual edges (e.g., metrics based on the so-called weak-ties hypothesis~\cite{Granovetter1973}, edge betweenness centrality~\cite{girvan2002}, or bridgeness~\cite{cheng2010}).
This is quite remarkable, as one might conjecture that an improved characterization of importance of edges in interaction networks could add to advance understanding and control of such networks~\cite{slotine2012}.

We here address this issue and propose modifications of various, widely used centrality concepts for vertices to those for edges, in order to find which edges in a network are important between other pairs of vertices.
We concentrate on eigenvector and closeness centrality, in addition to betweenness centrality~\cite{girvan2002}. 

Given the well known problem of ranking in networks~\cite{liao2017}, and in order to provide a more global (mesoscopic) view of importance of sets of edges and their role in structure and dynamics of interaction networks, we modify the concept of network decomposition to identify a bottom-up hierarchy of sets of edges, where each set is associated with a different level of importance.
We investigate characteristics of the proposed methods using paradigmatic network models and apply the novel concepts to identify important edges and important sets of edges in real-world networks.

\section{Estimating importance of vertices and edges}
\label{sec:centralities}

In the following, we consider undirected, binary or weighted networks that consist of sets of vertices $\mathcal{V}$ and edges $\mathcal{E}$, with $V = \left|\mathcal{V}\right|$ and $E = \left|\mathcal{E}\right|$ denoting the number of vertices and edges, respectively.

The importance of network constituents can be assessed with the concept of centrality that allows for various interpretations~\cite{lue2016}.
Centrality indices take into account the different roles network constituents play in a network. 
Here, we concentrate on the concepts of betweenness, closeness, and eigenvector centrality. 
Betweenness and closeness centrality are based on shortest paths, which requires the definition of ``length'' $d_{ij}$ of a path between vertices $i$ and $j$ or between edges $i$ and $j$.

The length $d_{ij}$ of a shortest path $P$ between vertices $i$ and $j$ in a binary network is the number of edges along this path.
We here utilize the same definition for the length $d_{ij}$ of $P$ between edges $i$ and $j$. 
For $i$ and $j$ being connected to a same vertex, we define $d_{ij} \defi 0$.
In case of a weighted network, we relate the length $d_{ij}$ of $P$ between vertices/edges $i$ and $j$ to the sum of the inverse weights of edges along this path~\cite{freeman1979}.
In case of adjacent edges, i.e., edges connected by a single vertex, we again define $d_{ij} \defi 0$.
 
\subsection*{Betweenness centrality}
\label{sec:betweennessDEF}
Following Refs.~\cite{newman2001c,barrat2004b,wang2008,opsahl2010,kuhnert2012}, vertex betweenness centrality (of vertex $k$) can be defined as
\begin{equation}
\mathcal{C}^{\rm B}_{\rm v}(k)=\frac{2}{(V-1)(V-2)}\sum_{k\neq i\neq j}\frac{q_{ij}(k)}{G_{ij}},
\label{eq:CBV}
\end{equation}
where $\left\{k,i,j\right\}\in\mathcal{V}$, and $q_{ij}(k)$ is the number of shortest paths between
vertices $i$ and $j$ running through vertex $k$. $G_{ij}$ is the total number
of shortest paths between vertices $i$ and $j$.

Similarly, edge betweenness centrality (of edge $k$) can be defined as~\cite{freeman1977,girvan2002}
\begin{equation}
\mathcal{C}^{\rm B}_{\rm e}(k)=\frac{2}{V(V-1)}\sum_{i\neq j}\frac{q_{ij}(k)}{G_{ij}}, 
\label{eq:CBE}
\end{equation}
where $k\in\mathcal{E}$, $\left\{i,j\right\}\in\mathcal{V}$, $q_{ij}(k)$ is the number of shortest paths between vertices $i$ and $j$ running through edge $k$, and $G_{ij}$ is the total number of shortest paths between vertices $i$ and $j$. 
A network constituent $k$ is the more important, the more commonly this constituent is part of the shortest path between every possible pair of vertices, excluding the pairs with vertex $k$.

\subsection*{Closeness centrality}
\label{sec:closenessDEF}
Vertex closeness centrality (for vertex $k$) is defined as 
\begin{equation}
\mathcal{C}^{\rm C}_{\rm v}(k)=\frac{V-1}{\sum_{l}^{}d_{kl}}
\label{eq:CCV}
\end{equation}
with $\left\{k,l\right\}\in\mathcal{V}$ and where $d_{kl}$ denotes the length of the shortest path from vertex $k$ to vertex $l$.

Similarly, we define edge closeness centrality (for edge $k$) as
\begin{equation}
\mathcal{C}^{\rm C}_{\rm e}(k)=\frac{E-1}{\sum_{l}^{}d_{kl}}, 
\label{eq:CCE}
\end{equation}
with $\left\{k,l\right\}\in\mathcal{E}$ and where $d_{kl}$ denotes the length of the shortest path from edge $k$ to edge $l$.
A network constituent $k$ is the more important, the shorter the paths that connect this constituent to every other reachable constituent of the same type. 

\subsection*{Eigenvector centrality}
\label{sec:eigenvectorDEF}

Vertex eigenvector centrality (of vertex $k$) is defined as the $k$th entry of
the eigenvector~$\vec{v}$ corresponding to the dominant eigenvalue
$\lambda_{\max}$ of matrix ${\bf M}$, which we derive from the eigenvector equation ${\bf M}\vec{v}=\lambda\vec{v}$ using the power iteration method:
\begin{equation}
\mathcal{C}^{\rm E}_{\rm v}(k)=\frac{1}{\lambda_{\max}}\sum_{l}^{}M_{kl}\,\mathcal{C}^{\rm E}_{\rm v}(l),
\label{eq:CEV}
\end{equation} 
with $\left\{k,l\right\}\in\mathcal{V}$. 
Here ${\bf M}$ denotes the adjacency matrix ${\bf A}^{\rm (v)} \in \left\{0,1\right\}^{V \times V}$ of a binary network, with $A^{\rm (v)}_{ij}=1$ if there is an edge between vertices $i$ and $j$, and 0 otherwise.
In case of a weighted network, ${\bf M}$ denotes the weight matrix ${\bf W}^{\rm (v)} \in \mathbb{R}_+^{V \times V}$, with $W^{\rm (v)}_{ij}$ denoting the weight of an edge between vertices $i$ and $j$. 
In a binary network, the degree~$k_i$ of a vertex~$i$ is defined as the number of its neighbors ($k_i \defi \sum_{j} A^{\rm (v)}_{ij}$). 
Its weighted counterpart is the strength $s_i \defi \sum_{j} W^{\rm (v)}_{ij}$.
We define $A^{\rm (v)}_{ii} \defi 0 \,\forall\, i$ and $W^{\rm (v)}_{ii} \defi 0 \,\forall\, i$ with $i\in\left\{1,\ldots,V\right\}$.

Analogously, we define edge eigenvector centrality (of edge $k$):
\begin{equation}
\mathcal{C}^{\rm E}_{\rm e}(k)=\frac{1}{\lambda_{\max}}\sum_{l}^{}M_{kl}\,\mathcal{C}^{\rm E}_{\rm e}(l),
\label{eq:CEE}
\end{equation} 
with $\left\{k,l\right\}\in\mathcal{E}$. 
Here ${\bf M}$ denotes the edge adjacency matrix ${\bf A}^{\rm (e)} \in \left\{0,1\right\}^{E \times E}$ of a binary network, with $A^{\rm (e)}_{ij}=1$ if edges $i$ and $j$ are connected to a same vertex, and 0 otherwise. 
In case of a weighted network, ${\bf M}$ denotes the weight matrix ${\bf W}^{\rm (e)} \in \mathbb{R}_+^{E \times E}$ whose entries $W^{\rm (e)}_{ij}$ are assigned the average weight of edges $i$ and $j$ if these edges are connected to a same vertex, and 0 otherwise. As above, we define $A^{\rm (e)}_{ii} \defi 0 \,\forall\, i$ and $W^{\rm (e)}_{ii} \defi 0 \,\forall\, i$ with $i\in\left\{1,\ldots,E\right\}$.
A network constituent $k$ is important if its adjacent constituents of the same type are also important.

With the aforementioned definitions, we regard a network constituent with the highest centrality value as most important and the one with the lowest centrality value as least important. 
In case of equal centrality values we rank in order of appearance.
This ranking allows us to compare findings achieved with the various centrality concepts.

\subsection*{Edge-centrality-based network decomposition}
\label{sec:decompose}
\begin{figure}[htbp]
	\includegraphics[width=0.4\textwidth]{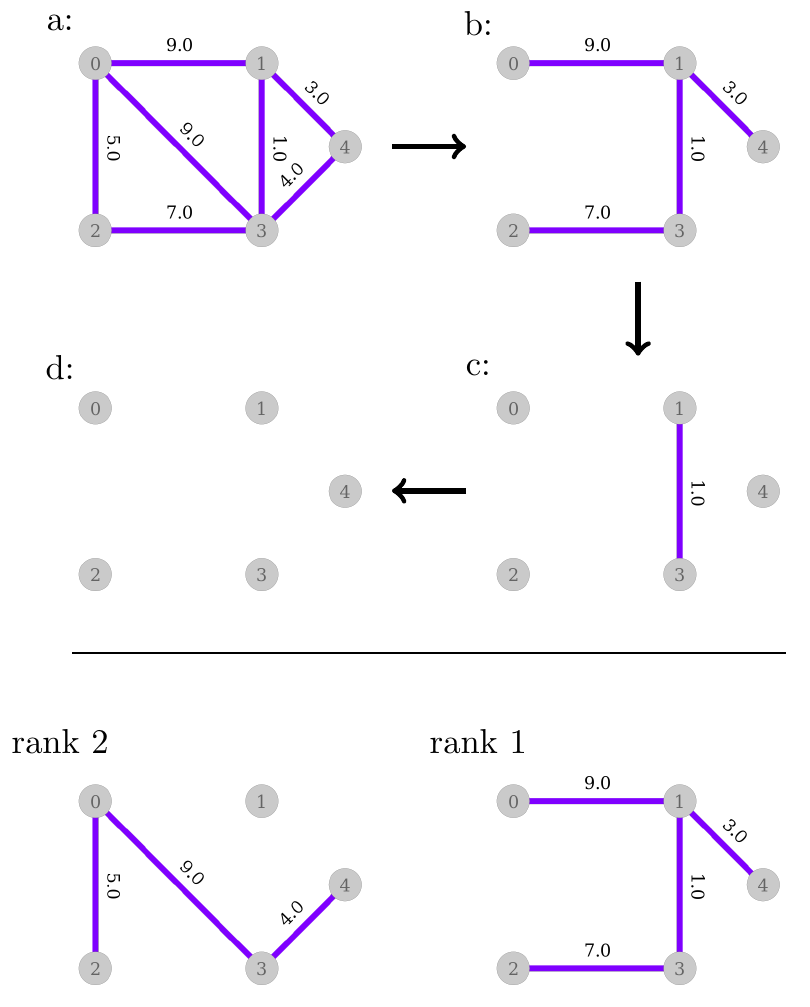}
	\caption{\CBE-based network decomposition of an example network (a) consisting of $V=5$ vertices and $E=7$ edges (numbers on edges denote edge weights).
	The current networks after the first step 3 in iteration 1, after the first step 3 in iteration 2, and after the second step 3 in iteration 2 are shown in (b-d). The resulting webs are shown in the lower part of the figure.}
	\label{fig:fig1}
\end{figure}

Network decomposition techniques such as the $k$-core/$k$-shell method~\cite{Kitsak2010} --~together with recent generalizations to weighted networks~\cite{Garas2012,Eidsaa2013,Eidsaa2016}~-- partition a network into sub-structures that are directly linked to vertex centrality. 
This allows to uncover tightly connected regions in a network such as sets of vertices with high centrality connected to other such vertices.
We here modify this concept and propose an edge-centrality-based network decomposition technique 
to identify a bottom-up hierarchy of sets of edges, where each set is associated with a different level of importance.
With an eye to a spider web, we denote the resulting set(s) of edges as ``web(s)''. 
Our technique is related to the $k$-core/$s$-core network decomposition~\cite{Eidsaa2013,Eidsaa2016} as well as to community detection algorithms~\cite{fortunato2010} and consists of the following steps to decompose a network with $E$ edges:
\begin{enumerate}[start=0]
\item initialize algorithm: set $E'=E$ and set iteration $n=1$;
\item estimate centrality $\mathcal{C}^{\bullet}_{\rm e}(m)$ for all edges $m\in \left\{1,\ldots,E'\right\}$ in the current network (where $\bullet \in \left\{\text{B,C,E}\right\}$);
\item choose the lowest centrality value as threshold value $\Theta=\min_m\mathcal{C}^{\bullet}_{\rm e}(m)$, in order to eliminate less central edges;
\item every edge $m'$ with $\mathcal{C}^{\bullet}_{\rm e}(m')\leq\Theta$ is assigned to the web of rank $n$ and is removed from the current network (which decreases $E'$); note that the $<$ sign holds for repetitions of step 3 within the $n$th iteration;
\item repeat step 1 and step 3 until no further edge is assigned to the web of rank $n$;
\item continue with next iteration (increase $n$ by 1) at step 1, as long as there are remaining edges to be assigned to webs;
\item reverse ranking of webs; the most important web has rank 1. 
\end{enumerate}
We note that the edge-centrality-based network decomposition can lead to two divisions of a network that are not helpful in identifying sets of edges associated with different levels of importance.
These cases are either an assignment of all edges to only one web (number of webs $\NWebs = 1$) or an assignment of each edge to a web ($\NWebs = E$). We also note that edges in a web do not have to be connected with each other.
\\
\\
We illustrate the algorithm by decomposing an example network (cf. Fig.~\ref{fig:fig1}a).
Starting with iteration $n=1$, \CBE-values is initially calculated in step 1 (estimated values at each step of iteration $n$ are summarized in Table~\ref{tab:tab1}). 
In step 2, the lowest centrality value is chosen as threshold ($\Theta=0.1$ marked red in Table~\ref{tab:tab1}).
In step 3, every edge with \CBE$=\Theta$ is assigned to the web with rank $n=1$ and removed from the current network. 
Note that there is no edge with \CBE$<\Theta$, since step 3 is executed in this iteration for the first time.
In step 4, the algorithm loops back to step 1 and \CBE of the remaining edges in the current network are calculated.
Step 2 is skipped and in step 3 no further edges are assigned to the web of rank $n=1$, since the newly calculated \CBE-values are larger than $\Theta$. 
Hence the condition to exit the loop in step 4 is achieved, and the next iteration $n=2$ starts.
In step 1, \CBE-values do not change, and in step 2 the new threshold $\Theta=0.4$ is chosen for this iteration.
In step 3, all remaining edges but one are assigned to the web of rank $n=2$.
In step 4, we enter the loop in this iteration, going back to step 1, newly calculating the centrality value of the remaining edge.
As only this edge is left in the network, one might expect \CBE$=1$ for this edge.
However, the decomposition algorithm only removes edges from the network, leaving the number of vertices unchanged (i.e. $V=5$ and not $V=2$ as one might expect for a network consisting of one edge only), thus yielding \CBE$=0.1<0.4=\Theta$.
Hence this remaining edge is also assigned to the web of rank $n=2$.
Lastly the ranking of webs is reversed and the algorithm yields two webs (see lower part of Fig.~\ref{fig:fig1}). 

\begin{table}[h]
	\begin{tabular*}{0.45\textwidth}{cccccccccc}
		\hline
		\hline
	  &      &      & \multicolumn{7}{c}{edge}                              \\
	  &      &      & (0,1) & (0,2) & (0,3) & (1,3) & (1,4) & (2,3) & (3,4) \\
	  \hline
	  n & step & ref. &       &       &       &       &       &       &       \\
	  1 & 1    & a    & 0.2   & 0.1   & 0.1   & 0.3   & 0.3   & 0.3   & 0.1   \\
	  & 2    & a   & 0.2   & \color{red}0.1   & \color{red}0.1   & 0.3   & 0.3   & 0.3   & \color{red}0.1   \\
	  & 3    & b    & 0.2   & -     & -     & 0.3   & 0.3   & 0.3   & -     \\
	  & 1*    & b    & 0.4   & -     & -     & 0.6   & 0.4   & 0.4   & -     \\
	  & 3*    & b    & 0.4   & -     & -     & 0.6   & 0.4   & 0.4   & -     \\
	  \hline
	  2 & 1    & b    & 0.4   & -     & -     & 0.6   & 0.4   & 0.4   & -     \\
	  & 2    & b    & \color{red}0.4   & -     & -     & 0.6   & \color{red}0.4   & \color{red}0.4   & -     \\
	  & 3    & c    & -     & -     & -     & 0.6   & -     & -     & -     \\
	  & 1*    & c    & -     & -     & -     & 0.1   & -     & -     & -     \\
	  & 3*    & d    & -     & -     & -     & -     & -     & -     & -    \\
		
		\hline
		\hline        
	\end{tabular*}
	\caption{Estimated \CBE-values at the respective decomposition steps during iteration $n$ of the example network shown in Fig.~\ref{fig:fig1}. Repeating steps during iteration $n$ and resulting from step 4 are marked with *. 
	\CBE-values chosen as threshold $\Theta$ are shown in red.
	Column `ref.' refers to subplots (a-d) in Fig.~\ref{fig:fig1}.}
	\label{tab:tab1}
\end{table}

\section{Characteristics of importance estimators}
For our investigations, we consider paradigmatic network models, namely small-world networks~\cite{watts1998} (with rewiring probability $p=0.1$, unless specified otherwise), scale-free networks~\cite{albert2002}, and random networks~\cite{erdos1959,batagelj2005}, 
for each of which we generated 1000 realizations.
For each realization of a weighted network, weights were drawn from the uniform distribution $\mathcal{U}(0,1)$. 
Other parameters for network generation were chosen such that networks consisted of $E=200$ edges, on average (small-world networks: four nearest neighbors; scale-free networks: each new vertex is attached to two existing ones; random networks: probability for edge creation equals edge density $\epsilon=2E(V(V-1)^{-1})$.

Here, we report our observations for weighted networks with $V=100$ vertices.
In general, findings obtained for binary networks are quite similar, and in the following we report on differences only.
We also obtained qualitatively similar findings for larger network sizes ($V=200$, $E\simeq400$) and thus expect that our findings carry over to even larger network sizes. 
We note though that computation times grows exponentially when increasing the number of edges $E$. 
\begin{figure*}[htp]
\begin{center}
	\includegraphics[width=\textwidth]{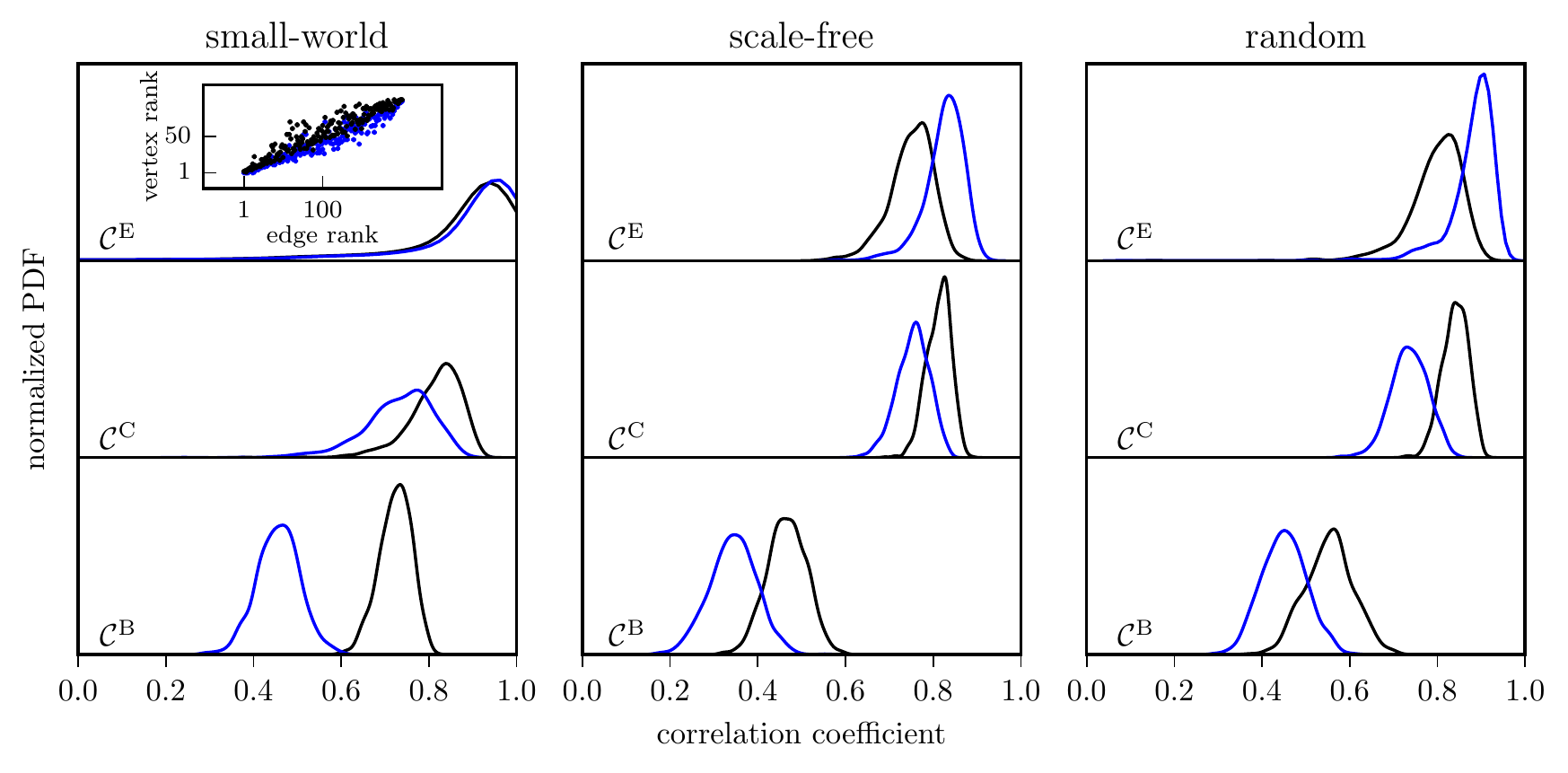}
\end{center}
\caption{Distributions of correlations between edge ranks and vertex ranks for the investigated topologies of weighted networks and for the various centrality concepts used to rate importance (top to bottom: \CE, \CC, \CB). 
Data are shown separately for the two vertices connected by an edge: blue colors refers to the vertex with higher importance (\vplus) and black colors to the less important partner (\vminus). 
The inset shows an exemplary scatterplot of vertex ranks versus edge ranks for one realization of a small-world network.}
	\label{fig:fig2}
\end{figure*}
\subsection{Do important edges connect important vertices?}
Due to the analogous definition of a given centrality for vertices and edges, one might expect that important edges connect important vertices.
For example, with betweenness centrality as an estimate for importance, one might deduce that if the flow of information through some edge is high, the flow through the vertices connected by that edge is high as well, and an analogous reasoning can be made for closeness centrality. 
With eigenvector centrality, a vertex is the more important the more important its neighboring vertices are; thus, an edge connecting two such vertices can also expected to be important.
It has to be taken into account, however, that more than one edge can be linked to a single vertex and that in a connected network the number of edges is generally higher than the number of vertices.

In order to answer the question whether important edges connect important vertices, we rank --~for a given centrality concept~-- importance of vertices and edges by assigning the vertex/edge with the highest centrality value rank 1, the second-highest centrality value rank 2, etc. (in case of equal centrality values we rank in order of appearance).
In the following, we denote with (\vplus,\vminus) a pair of vertices, connected by an edge, with vertex \vplus having a higher importance than its partner \vminus.
We perform correlation analyses of edge ranks and ranks of vertices \vplus and of \vminus.

Fig.~\ref{fig:fig2} summarizes our findings for weighted networks.
With eigenvector centrality, we observe for all topologies high correlation values (Pearson's $r>0.8$ on average) between edge ranks and ranks of \vplus vertices but with a higher abundance for scale-free and random networks. 
For these topologies, correlations between edge ranks and ranks of \vminus vertices are less pronounced and less often, in contrast to small-world networks for which distributions are quite similar.
With closeness and betweenness centrality, correlation values decrease in general (and even reach mean values less than 0.5), nevertheless we observe with both centralities higher correlations between edge ranks and ranks of \vminus vertices than with ranks of \vplus vertices.
These findings point to a linear relationship between edge and vertex ranks in these binary and weighted networks.

Given the high correlations between edge ranks and vertex ranks, we investigate --~for a given centrality concept~-- the frequency for an edge with highest importance (maximum edge centrality) to link to a vertex with highest importance (maximum vertex centrality). Furthermore we investigate the median importance rank of the vertex connected via that edge to the vertex with highest importance.
\begin{table}[h]
\begin{tabular*}{0.4\textwidth}{@{\extracolsep{\fill} } *{7}{c}}
\hline
\hline
\multicolumn{1}{l}{weighted}  & \multicolumn{2}{c}{small-world} & \multicolumn{2}{c}{scale-free} & \multicolumn{2}{c}{random} \\ 
\cline{2-3}
\cline{4-5}
\cline{6-7}
\multicolumn{1}{c}{network}  & \PER & \MEDR & \PER & \MEDR & \PER & \MEDR \\ 
\hline
\CE \qquad & 68 & 3 & 90 & 3 & 76 & 3 \\
\CC \qquad & 55 & 4 & 88 & 3 & 71 & 3 \\
\CB \qquad & 67 & 3 & 77 & 3 & 59 & 4 \\ 
\hline\hline
\end{tabular*}
\caption{Percentage (\PER) of the weighted small-world, scale-free, and random networks, for which an edge with highest importance (as identified with maximum edge centrality (\CEE, \CCE, or \CBE)) is connected to a vertex with highest importance (as identified with maximum vertex centrality (\CEV, \CCV, or \CBV)). 
The median rank of the opposing vertex is denoted by \MEDR.}
\label{tab:tab2}
\end{table}
Depending on the centrality concept (see Table~\ref{tab:tab2}), we observe such highest-importance connections in about two-thirds of our realizations of small-world and random networks and in an even higher percentage (up to 90\,\%) of scale-free networks.
The importance of the opposing vertex has rank 3, on average.
With these findings, we can conclude that important edges indeed connect important vertices in a large majority of the investigated networks and that the three centrality concepts rate importance of edges differently, as expected.

\subsection{Do important edges highlight typical properties of network topologies?}
\begin{figure}[htp]
\begin{center}
 \includegraphics[width=.5\textwidth]{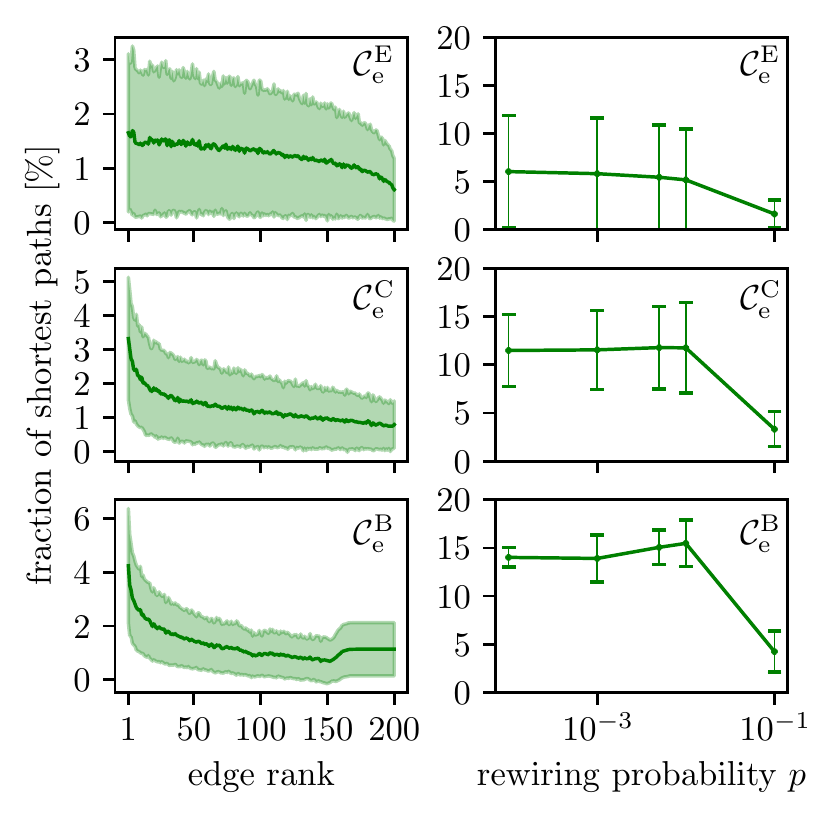}
\end{center}
\caption{Left: Fraction of shortest paths in weighted small-world networks (rewiring probability $p=0.1$) that traverse an edge of a given rank.
Edge ranks estimated with eigenvector centrality (top), closeness centrality (middle), and betweenness centrality (bottom).
Means and standard deviations (green lines and light-green shaded areas) from 1000 realizations of networks with 200 edges each.
Right: Dependence of fraction of shortest paths traversing the most important edge on rewiring probability $p$.
 Lines are for eye guidance only.}
	\label{fig:fig3}
\end{figure}
\begin{figure}[htp]
\begin{center}
	\includegraphics[width=.5\textwidth]{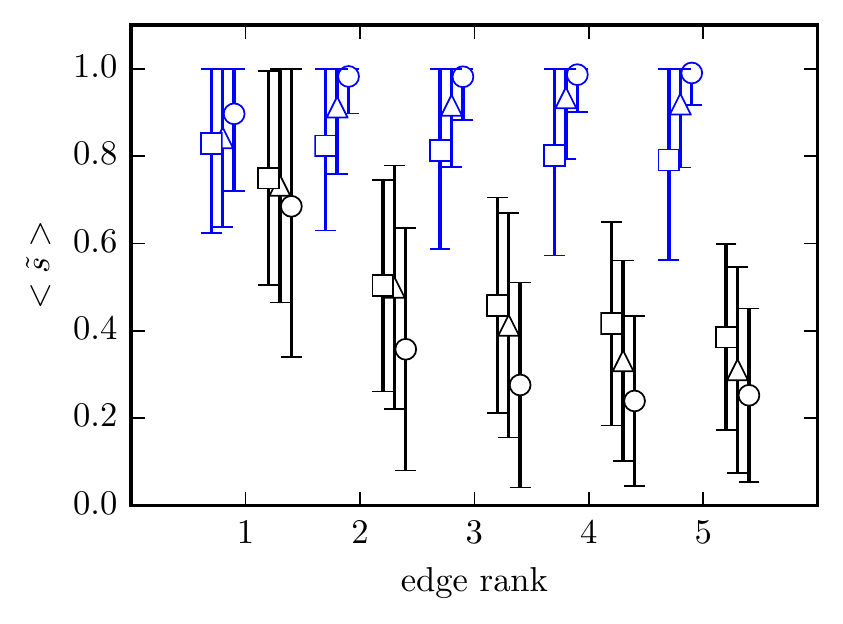}
\end{center}
\caption{Means and standard deviations of normalized strength \rmstrength of \vplus vertices (blue) and \vminus vertices (black) connected by the five most important edges in 1000 weighted scale-free networks. 
Edge importance estimated with eigenvector centrality (\CEE, circles), closeness centrality (\CCE, triangles), and betweenness centrality (\CBE, squares).}
	\label{fig:fig4}
\end{figure}
Addressing this question, we investigate in small-world networks a possible relationship between edge centrality and shortest paths, and in scale-free networks which vertices most important edges typically connect (e.g., core-core, core-periphery, or periphery-periphery vertices).

For our small-world networks, we observe that edges with highest rank are, on average, part of most shortest paths compared to edges with lower rank (cf. Fig.~\ref{fig:fig3}).
As long-range connections in small-world networks are also part of most shortest paths by construction, this indicates that long-range connections are the most important edges in these networks, if importance was assessed with closeness or betweenness centrality, and to a lesser degree also with eigenvector centrality.

For our weighted scale-free networks, we show in Fig.~\ref{fig:fig4} the normalized strength \rmstrength $\left(\rmstrength=s(k)/\max_{v \in \mathcal{V}} s(v); k \in\left\{\vplus,\vminus\right\}\right)$ of \vplus and \vminus vertices connected by the five most important edges (rank 1~--~5). 
\vplus vertices that belong to the edge with highest importance (maximum centrality value) have a high relative mean strength ($\rmstrength > 0.8$) while that of the opposing \vminus vertex is about 0.1 smaller. 
For eigenvector and closeness centrality, we observe highest \rmstrength values for \vplus vertices belonging to edges with importance ranks between 2 and 24, resp, 2 and 14.
These findings indicate that high-importance edges typically connect vertices within or close to the core of a weighted scale-free network. 

\subsection{Can different edge centrality concepts identify the same most important edge?}
\begin{figure}[htp]
\begin{center}
	\includegraphics[width=0.5\textwidth]{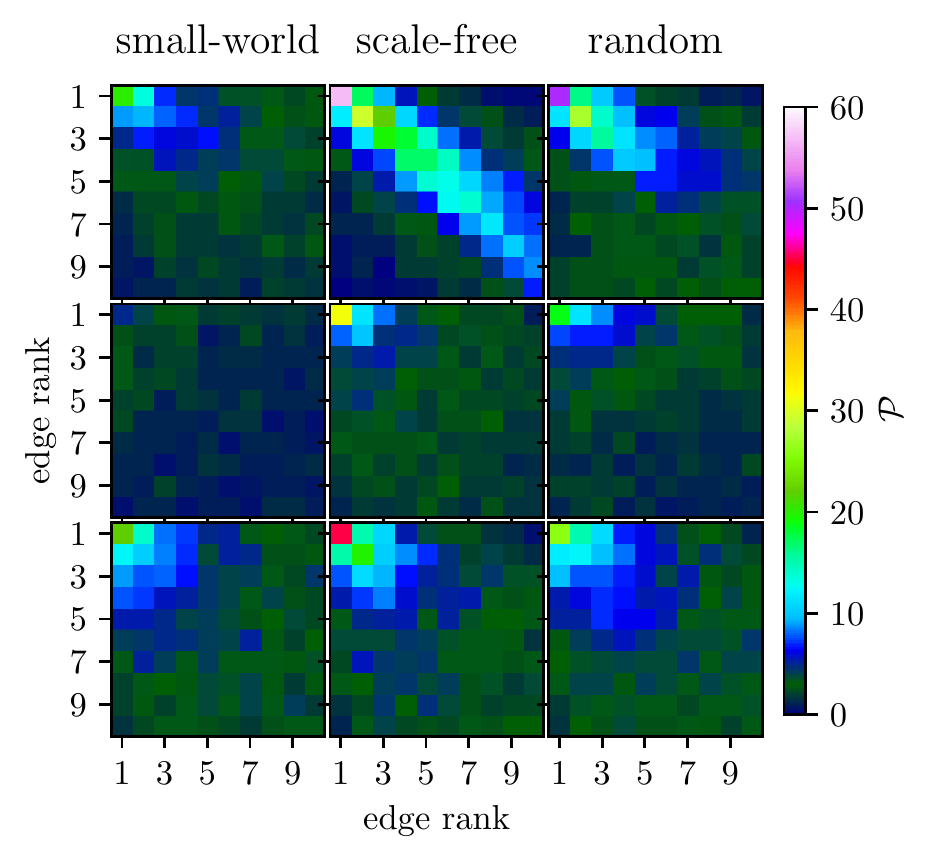}
\end{center}
\caption{Percentage $\cal{P}$ of weighted small-world (left), scale-free (middle), and random networks (right) with 1000 realizations each for which different edge centralities (\CEE, \CCE, \CBE) identify the same edge with a given edge rank (top: \CEE vs. \CCE, middle: \CEE vs. \CBE, bottom: \CCE vs. \CBE).}
	\label{fig:fig5}
\end{figure}
Previous studies~\cite{Bolland1988,Stephenson1989,Nakao1990,Rothenberg1995,Borgatti2005,Newman2005,Manimaran2009,Valente2008,Nguyen2011,Wang2011,kuhnert2012} reported on correlations between various vertex centralities, which can lead to identifying the same vertex as most important using different centrality concepts.
Although such a finding might underline the high importance of that vertex, its  interpretation is nonetheless hampered.

Following this line of research, we estimate the frequency of deriving the same edge rank with different edge centrality concepts.
Fig.~\ref{fig:fig5} summarizes our findings for all investigated weighted networks.
We find with eigenvector and closeness edge centrality the same most important edge in about 50\,\% of our scale-free and random networks, and in about 20\,\% of our small-world networks. 
Similarly, closeness and betweenness edge centrality identify the same most important edge in 45\,\% of our scale-free networks. 
For other networks and combinations of edge centrality concepts, we obtain concordant findings in less than 30\,\% of cases and for most edges with a rank larger than or equal to two, concordance only rarely exceeds chance level.

In case of binary networks, we find with eigenvector and closeness edge centrality the same most important edge in about 63\,\%, 79\,\%, and 50\,\% respectively in our small-world, scale-free, and random networks.
For other networks and combinations of edge centrality concepts, we obtain concordant findings in less than 40\,\% of cases.

The high concordance levels seen for weighted scale-free and random networks follow from the chosen weight distribution. 
With weights drawn from a uniform distribution, there is a high probability for edges with highest weights to be connected to high-strength (high-degree) vertices.
Such a configuration is predestined for an identification of the most important edge with eigenvector and closeness edge centrality.
We note that other weight distributions can lead to different findings.

\begin{figure}[htp]
	\begin{center}
		\includegraphics[width=.49\textwidth]{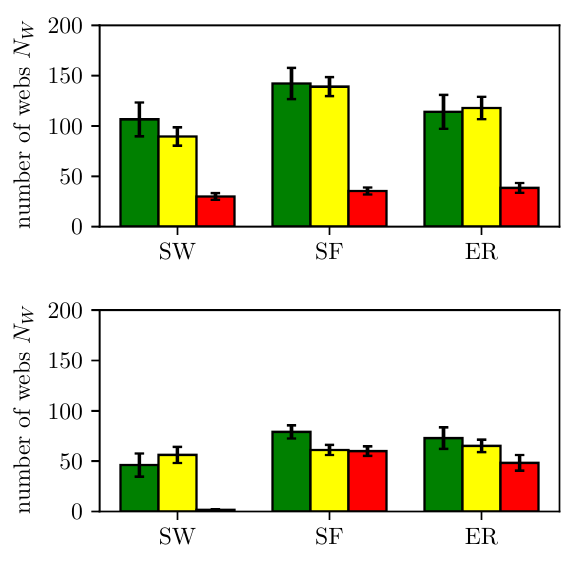}
	\end{center}
	\caption{
		Number of webs \NWebs (means and standard deviations) derived with the edge-centrality-based network decomposition technique using different edge centralities (\CEE: green, \CCE: yellow, \CBE: red). 
		Data obtained from 1000 realizations of weighted (top) and binary (bottom) networks consisting of 200 edges.}
	\label{fig:fig6}
\end{figure}

\subsection{Substructures in important webs}

Given that edges in a web do not have to be connected with each other, we now define a substructure to be connected and consist of at least two edges or to be a solitary edge and investigate the substructure's characteristics. 

We begin by determining the number of webs \NWebs that can be achieved with a given edge centrality concept for the investigated weighted network topologies, with $E=200$ edges each (see Fig.~\ref{fig:fig6}).
This number, which is a priori not known, allows us to rule out the above-mentioned cases of \NWebs$=1$ (web containing one substructure consisting of all edges) or \NWebs~$=E$ (webs containing one substructure being a solitary edge).
A decomposition based on eigenvector or closeness edge centrality leads to about 100 webs for weighted small-world and random networks and to about 140 webs for weighted scale-free networks, on average. 
In contrast, if the decomposition was based on betweenness edge centrality, we achieve about 25 webs, on average, for all investigated weighted network topologies. 
Depending on initial conditions, we observe highest spread for the eigenvector-edge-centrality-based decomposition, followed by the closeness-edge-centrality-based decomposition, and the smallest one for the betweenness-edge-centrality-based decomposition.
Apart from \NWebs being reduced by a factor of about 2, we obtain comparable results for binary networks, with the exception of the \CBE-based decomposition, for which \NWebs varies depending on topology. 
These findings indicate that the number of achievable webs depends on the network topology and to a greater extent on the edge centrality used for the decomposition. 
We note that other weight distributions can lead to different findings.

\begin{figure*}[htb]
	\begin{center}
		\includegraphics[width=1.\textwidth]{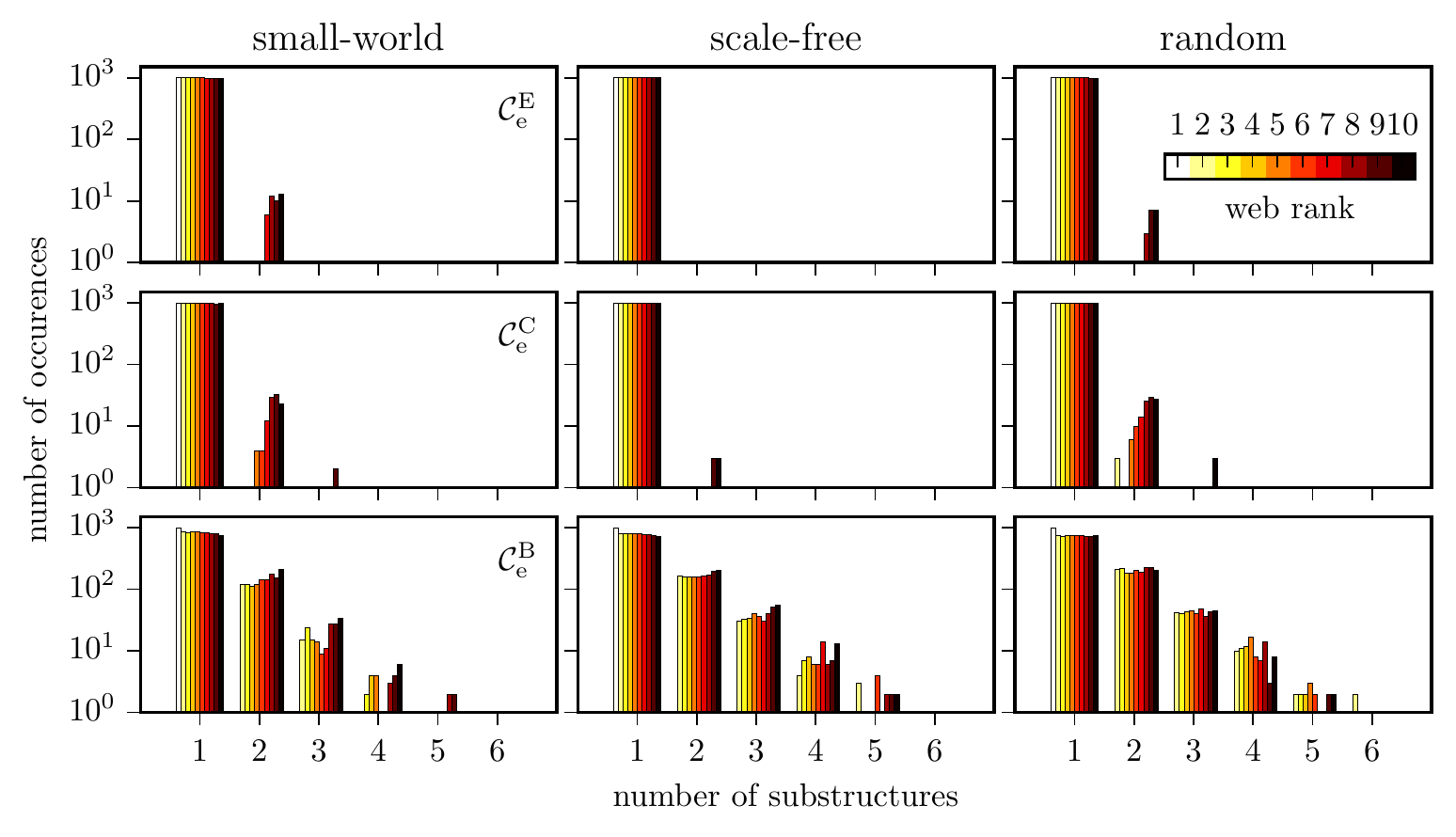}
	\end{center}
	\caption{Occurrences of a certain number of substructures within the 10 most important webs (color coded) in 1000 realizations of weighted small-world (left), scale-free (middle), and random networks (right). 
		Edge-centrality-based network decompositions were performed with different edge centralities (from top to bottom: \CEE, \CCE, \CBE).}
	\label{fig:fig7}
\end{figure*}
\begin{figure*}[htp]
	\begin{center}
		\includegraphics[width=1\textwidth]{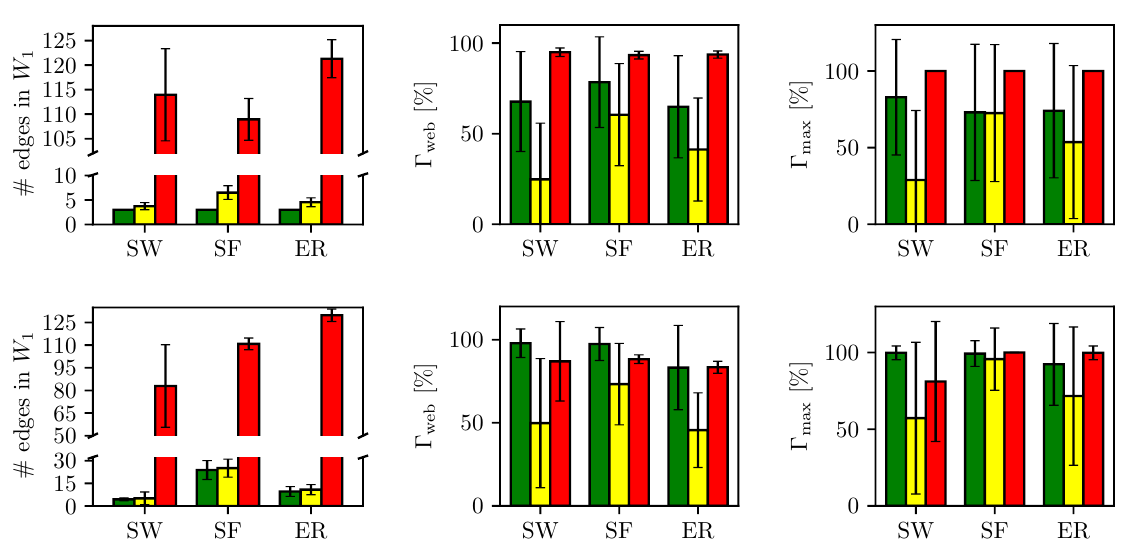}
	\end{center}
	\caption{Left: Number of edges merged into the most important web $W_1$. Middle: Percentage overlap $\Gamma_{\mathrm{web}}$ of edges in $W_1$ and $n_\mathrm{e}$ most important edges, with $n_\mathrm{e}$ being the number of edges in $W_1$ (see left). Right: Percentage $\Gamma_{\mathrm{max}}$ of most important webs $W_1$ containing the most important edge.
		Edge-centrality-based network decompositions performed with different edge centralities (\CEE: green, \CCE: yellow, \CBE: red). 
		Means and standard deviations obtained from 1000 realizations of weighted (top) and binary (bottom) small-world (SW), scale-free (SF), and random (ER) networks, each consisting of 200 edges.}
	\label{fig:fig8}
\end{figure*}

Having identified the number \NWebs of webs, we next investigate the number of substructures important webs are composed of.
Fig.~\ref{fig:fig7} summarizes our findings for the ten most important webs in weighted networks.
For each topology, and independent of the employed edge centrality concept, we observe the most important web $W_1$ (web with rank 1) to be composed of one substructure.
Depending on the employed edge centrality concept for the decomposition, this single substructure consists of three edges in small-world and of, on average, five edges in scale-free as well as of 115 edges in random weighted networks (cf. Fig.~\ref{fig:fig8}, upper left; note that the zero variance of the number of edges in $W_1$ seen with the \CEE-based decomposition for all weighted network topologies is related to the concept underlying this edge centrality).
For our binary networks, we mostly obtain comparable findings with the number of edges in the single substructure being of the same order as seen with weighted networks (cf. Fig.~\ref{fig:fig8}, lower left).
An exception is the two- to eight-fold higher number of edges in $W_1$ of scale-free and random binary networks achieved with the \CEE-~and \CCE-based decompositions.
This can be related to the occurrence of recurrent, identical patterns (being higher in scale-free than in random networks) which is well captured by these edge centralities.
The in general higher number of edges in $W_1$ as yielded by the \CBE-based network decomposition can be related to a discretization effect: this edge centrality only takes into account the number of shortest paths and hence does not directly depend on the numerical value of the respective path lengths.

We note that edges, which have been assigned to the most important \CEE-based and \CCE-based webs, connect vertices of which more than two-thirds have a degree of one (data not shown).
This indicates that substructures in these webs are star-like, since the majority of their vertices are adjacent to only one of a few high-degree vertices that represent the center of these star-like substructures. This holds for the topologies investigated here.

As regards the most important \CBE-based web, the number of vertices with degree one highly depends on the topology of the network. For small-world, scale-free, and random networks we find about $30\%$, $60\%$ and $40\%$, respectively, of vertices having degree one. Star-like substructures can thus be observed preferentially in the most important webs of scale-free networks.

Having investigated characteristics of substructures in most important webs, we now address the question if these webs contain most important edges.
In Fig.~\ref{fig:fig8} (middle) we report the percentage overlap $\Gamma_\mathrm{web}$ of edges in most important webs, consisting of $n_\mathrm{e}$ edges, with the $n_\mathrm{e}$ most important edges. 
Independent of topology but depending on the employed centrality concept, we observe a rather small percentage overlap ($\Gamma_\mathrm{web}<75\%$) for \CCE and a rather large one ($\Gamma_\mathrm{web}>90\%$) for \CBE in both weighted and binary networks. For \CBE, this high overlap is to be expected, given the high number of edges in $W_1$.
For \CEE, $\Gamma_\mathrm{web}$ takes on comparably high values of up to $100\%$ in binary networks, while we find $\Gamma_\mathrm{web}<75\%$ in weighted networks.
Interestingly, we obtain similar results if we consider the most important edge only (Fig.~\ref{fig:fig8} (right)).
These findings clearly indicate that the most important edge as assessed with the different edge centralities is not necessarily contained in the most important web.

We conclude that webs derived with our edge-centrality-based network decomposition contain substructures of important edges, however not inevitably containing the most important edge, in the networks analyzed here. 

\section{Application to real-world networks}
\begin{figure}[htp]
\begin{center}
	\includegraphics[width=.5\textwidth]{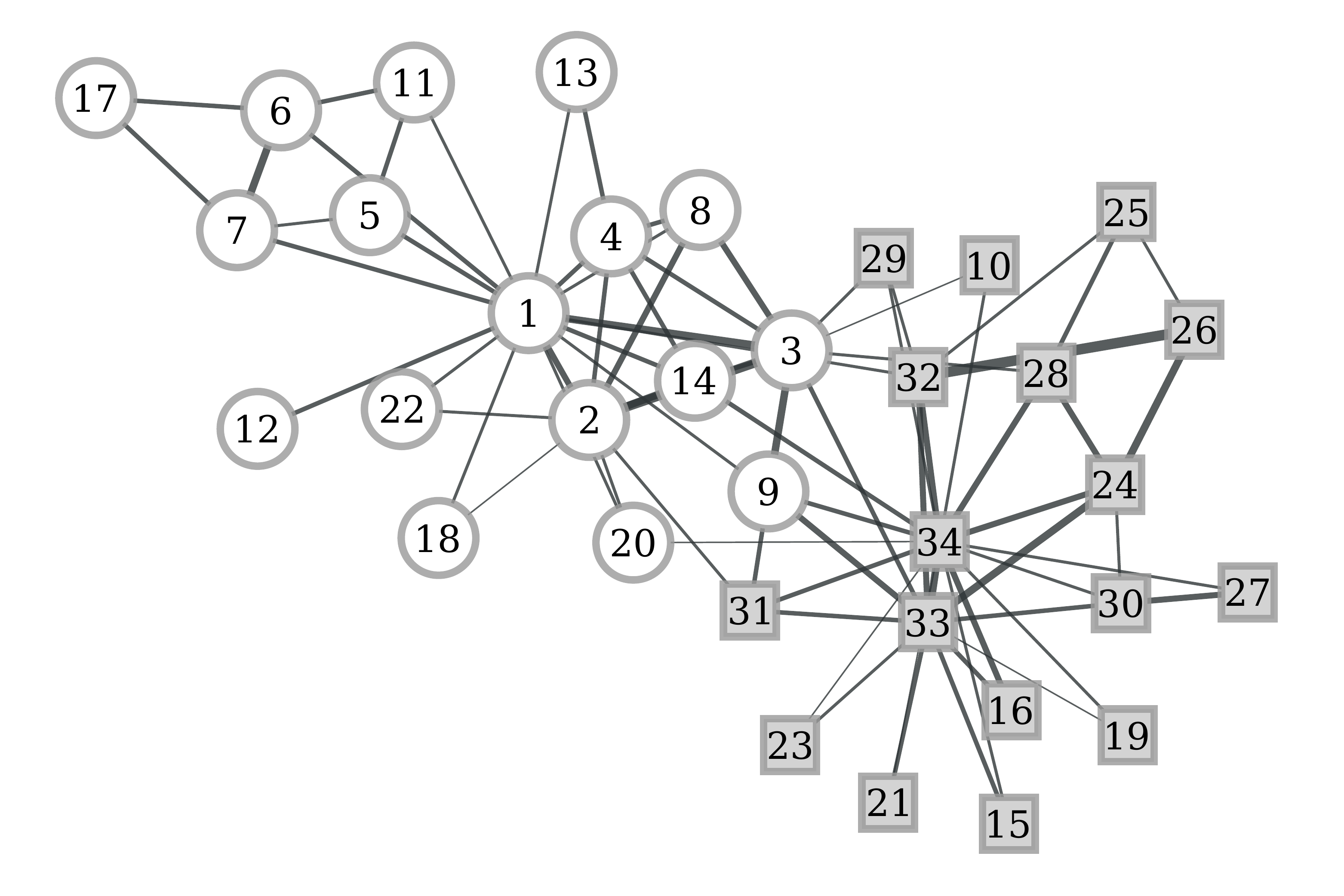}
\end{center}
\caption{Zachary's karate club network. 
Numbered vertices represent the members of the club and edges represent friendships, as determined by observation of interactions. 
Edge weights are encoded as line thickness.
The two factions into which the club split during the course of the study are indicated by circles (the club instructor's group (vertex 1)) and squares (the club administrator's group (vertex 34)).}
\label{fig:fig9}
\end{figure}
We now demonstrate the utility of our proposed methods for understanding which edges or sets of edges in a real-world network are important, by testing them on a commonly used benchmark model in social network analysis with known hierarchical structure, namely the karate club at a US university in the 1970s studied in detail by Zachary~\cite{zachary1977},
and on evolving functional brain networks from an epilepsy patients transiting into and out of an epileptic seizure~\cite{geier2015}.

\subsection{Zachary's karate club network}

The network consists of 34 persons (vertices) whose mutual friendship relationships (78 edges) have been carefully investigated over a period of two years. 
Due to contrasts between the club's instructor (vertex 1) 
and the administrator (vertex 34), 
the club finally split into two smaller groups (see Fig.~\ref{fig:fig9}), which ultimately resulted in the instructor's leaving and starting a new club, taking about a half of the original club's members with him. 

We begin by investigating the importance of edges and vertices using the various centrality concepts (see Sec.~\ref{sec:centralities}) and report our findings in Tab.~\ref{tab:tab3}.
For the weighted karate club network, we find with vertex closeness centrality vertex 34 (club administrator) as most important followed by vertex 1 (the club's instructor), and these vertices switch position in ranking when rating their importance with vertex betweenness centrality.
With vertex eigenvector centrality, vertices 1 and 34 have rank 4 and 3, respectively.  
For the binary karate club network, all vertex centralities identify either vertex 1 or vertex 34 as most important.
These findings confirm previous studies~\cite{everett1998,batool2014,qi2015,qiao2017}.

Interestingly, with all employed edge centralities we find the most important edge to connect pairs of vertices of which one partner is either vertex 1 or vertex 34.
Highest-ranked edges are identical for the binary and the weighted karate club network.
None of the edge centralities indexes a direct important connection between vertices 1 and 34, as expected.
We note though that edge betweenness centrality points to a shortest path between these vertices via vertex 32.
\begin{table}[h]
    \begin{tabular*}{0.4\textwidth}{@{\extracolsep{\fill} } *{6}{c}}
      \hline\hline
      & weighted & \multicolumn{4}{c}{rank}  \\
      \cline{3-6}
      & network & 1  & 2  & 3  & 4  \\
      \hline
      \multirow{2}{*}{\CE} & vertex & 3  & 33  & 34  & 1  \\
      & edge & \{33,34\} & \{34,32\}  & \{34,24\} & \{34,28\}\\  
	    \multirow{2}{*}{\CC} & vertex & 34  & 1  & 20  & 22  \\
      & edge & \{1,3\}  & \{34,33\}  & \{2,3\}  & \{9,3\}  \\
      \multirow{2}{*}{\CB} & vertex & 1  & 34  & 32  & 33  \\
      & edge & \{1,32\} & \{34,32\} & \{34,33\} & \{31,2\} \\
      \hline\hline
    \end{tabular*}
    \begin{tabular*}{0.4\textwidth}{@{\extracolsep{\fill} } *{6}{c}}
       	& binary & \multicolumn{4}{c}{rank}  \\
       	\cline{3-6}
       	& network & 1  & 2  & 3  & 4  \\
       	\hline
       	\multirow{2}{*}{\CE} & vertex & 34 & 1  & 3  & 33  \\
       	& edge & \{34,33\} & \{34,9\}  & \{34,32\} & \{33,14\}\\  
       	\multirow{2}{*}{\CC} & vertex & 1  & 3  & 34  & 32  \\
       	& edge & \{1,3\}  & \{1,32\}  & \{34,32\}  & \{3,32\}  \\
       	\multirow{2}{*}{\CB} & vertex & 1  & 34  & 33  & 3  \\
       	& edge & \{1,32\} & \{1,6\} & \{1,7\} & \{1,3\} \\
       	\hline\hline
    \end{tabular*}
    \caption{The top-4 ranked vertices and edges of the karate club network. Importance estimated with eigenvector centrality (\CE), closeness centrality (\CC), and betweenness centrality (\CB). For edges, we report the pair of vertices (\vplus, \vminus) connected by that edge.}
    \label{tab:tab3}
\end{table}

\begin{figure}[htp]
\begin{center}
	\includegraphics[width=.45\textwidth]{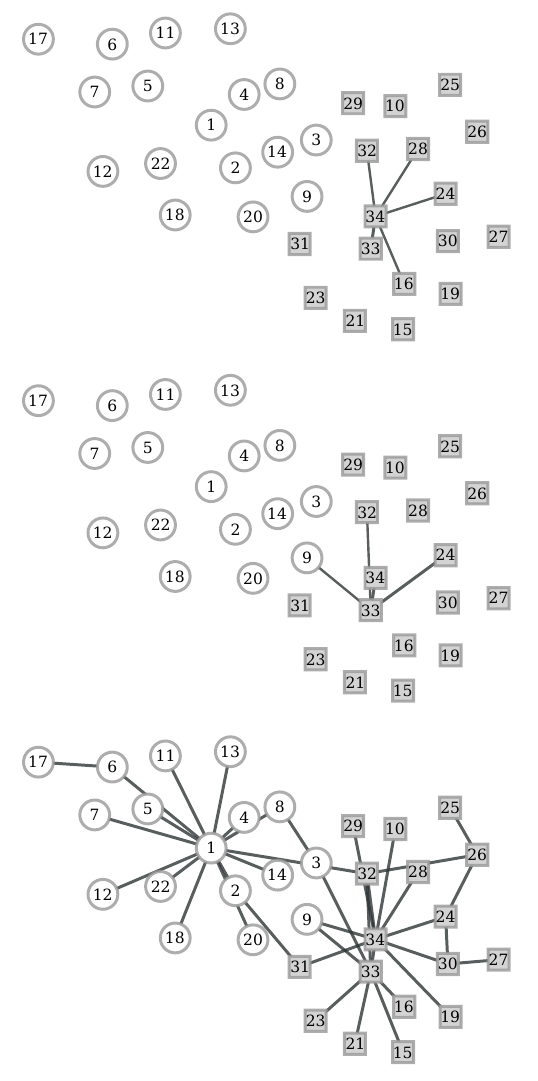}
\end{center}
\caption{Most important web (rank 1) obtained with an edge-centrality-based decomposition of the weighted karate club network using edge eigenvector (\CEE; top), edge closeness (\CCE; middle), and edge betweenness centrality (\CBE; bottom). 
The number of resulting webs amounted to $\NWebs=26$ using \CEE, $\NWebs=43$ using \CCE, and $\NWebs=20$ using \CBE.}
	\label{fig:fig10}
\end{figure}

We proceed with our edge-centrality-based network decomposition and show in Fig.~\ref{fig:fig10} the most important webs in the weighted karate club network obtained with eigenvector (\CEE), closeness (\CCE), and betweenness edge centrality (\CBE).
With a decomposition based on \CEE or on \CCE, the most important web (rank 1) takes on a star-like form with edges connecting the club administrator (vertex 34) to few other members of his faction (except for vertex 9 in case of \CCE). 
We observe a similar star-like web with edges connecting the club instructor (vertex 1) to many other members of his faction at rank 11 only using the \CEE-based decomposition (data not shown). 
No such web is observed with the \CCE-based decomposition.
The heterogeneous occurrences of star-like webs centered around either vertex 34 or vertex 1 can be explained by the different connections of these two vertices with the respective opposing fission group.
While vertex 1 is connected to 15 out of 16 members of his faction and to one member of the opposing faction, vertex 34 is connected to 14 out of 18 members of his faction and to three members of the opposing faction. 
The in general higher number of connections of vertex 34 explains the findings obtained with both the \CEE- and the \CCE-based  decomposition. 
Interestingly, the most important web obtained with the \CBE-based decomposition essentially takes on two star-like forms, both with edges connecting either the club instructor (vertex 1) or the club administrator (vertex 34) to almost all members of theirs factions.
We also find two paths in this web that connect vertices 1 and 34, namely via vertices 3 and 32 as well as via 2 and 31, and both these edges lie on shortest paths in the karate club network (cf. Fig.~\ref{fig:fig9}).
\begin{figure}[htp]
\begin{center}
	\includegraphics[width=.45\textwidth]{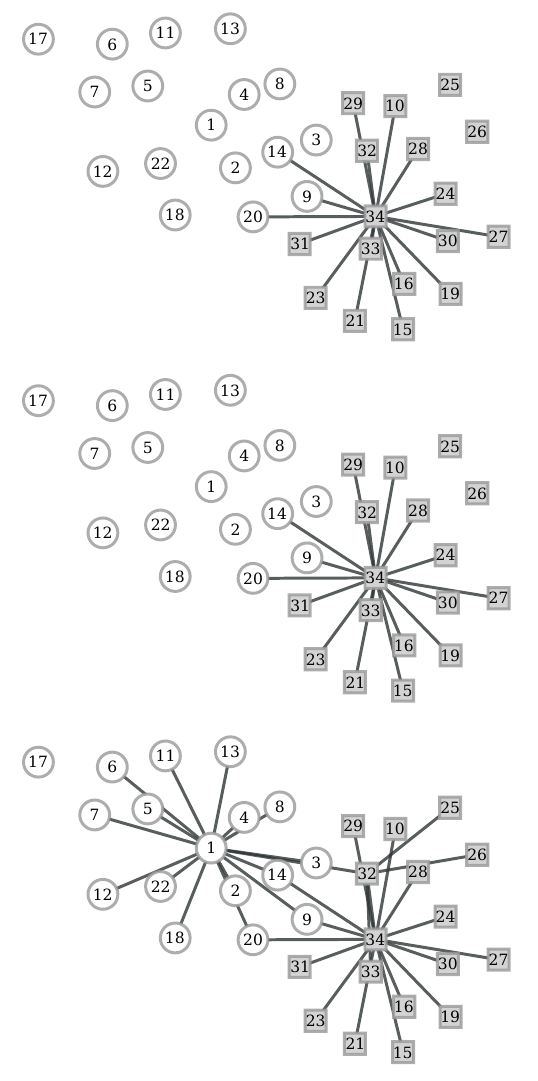}
\end{center}
\caption{Same as Fig.~\ref{fig:fig10} but for the binary karate club network.
The number of resulting webs amounted to $\NWebs=22$ using \CEE, $\NWebs=22$ using \CCE, and $\NWebs=24$ using \CBE.}
	\label{fig:fig11}
\end{figure}

If we consider the binary karate club network, the most important web obtained with the \CBE-based decomposition resembles to a large extent the one obtained for the weighted network.
The most important webs derived with the \CEE- and \CCE-based decompositions are identical and take on a star-like form with edges connecting the club administrator (vertex 34) to almost all other members of his faction and to a few members of the other faction.
With lower-rank webs (\CEE: rank 4; \CCE: rank 2), we find a similar star-like form but centered at the club instructor (vertex 1) with edges connecting to almost all members of this faction and to only one member of the other faction (data not shown).

We note that mismatched or unmatched vertices in the different decompositions are related to club members that had been identified by Zachary to only have a weak affiliation with either of the two groups.

Overall, our edge centralities identified most important edges in the karate club network that connect pairs of vertices of which one partner (either club instructor or club administrator) is known to be most important a priori (and confirmed to be most important with various vertex centralities). 
Our edge-centrality-based network decomposition revealed sets of edges that connect the two most important vertices (club instructor and club administrator) to their respective factions. 
Depending on the used edge centrality concept, this information is either contained in one web (typically the most important one) or spreads across multiple webs.

\subsection{Evolving functional brain networks during an epileptic seizure}
\begin{figure}[htp]
\begin{center}
\includegraphics[width=.5\textwidth]{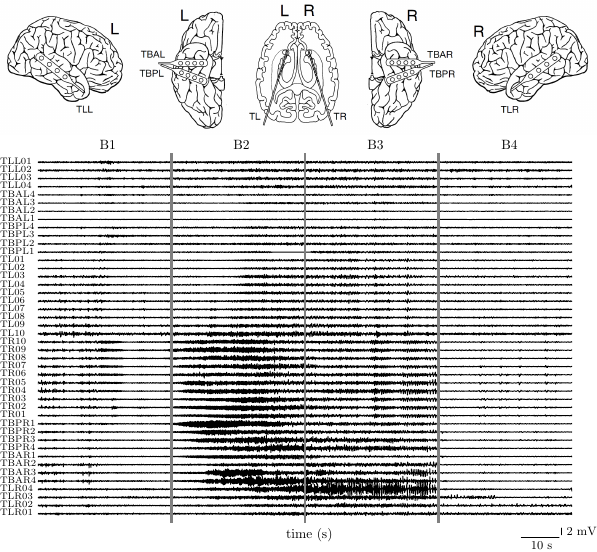}
\caption{
(A) Schematics of sensors placed over the left and right temporal-lateral and temporal-basal neocortex and of bilateral intrahippocampal sensors. 
(B) Invasive electroencephalographic recording of a seizure from the left (upper half) and right brain hemisphere (lower half). 
Block B1 indicates the pre-seizure phase; blocks B2 and B3 indicate the first and second half of the seizure (electroencephalographic seizure onset and ending determined automatically~\cite{schindler2007a}); block B4 indicates the post-seizure phase.}
\label{fig:fig12}
\end{center}
\end{figure}

As a second example, we investigate evolving functional brain networks~\cite{lehnertz2014} that we derived from multichannel electroencephalographic (EEG) data recorded from an epilepsy patient prior to, during, and after a focal-onset seizure (see Fig.~\ref{fig:fig12}). 
Vertices of such networks are usually associated with sensors that are placed to sufficiently capture the dynamics of the sampled brain region, and edges represent time-varying interactions between pairs of brain regions.
The data were part of previous studies~\cite{schindler2007a,schindler2008a,bialonski2011b,geier2015,geier2017a,stahn2017} and were recorded from sensors (chronically implanted strip and depth electrodes; see Fig.~\ref{fig:fig12}) placed on the cortex and within relevant brain structures during the presurgical evaluation of the patient's medically uncontrollable epilepsy. 
Decisions regarding sensor placement were purely clinically driven and were made independently of this study. 
EEG data from 44 sensors were sampled at \unit[200]{Hz}, digitized using a 16-bit analog-to-digital converter, filtered within a frequency range of \unit[0.1--70]{Hz}, and referenced against the average signal of two sensors outside the focal region.
The patient had signed informed consent that the clinical data might be used and published for research purposes. 
The study protocol had been approved by the ethics committee of the University of Bonn.
It is still matter of debate~\cite{zaveri2009,warren2010,wilke2011,varotto2012,geier2015,zubler2015,goodfellow2016,khambhati2016,li2016} whether the region of the epileptic brain from which first signs of seizure activity were recorded --~the so called seizure onset zone (SOZ)~-- is the most important network constituent.
Addressing this issue, many previous studies employed vertex centralities, however, an investigation based on edge centralities has not yet been carried out.
Furthermore, characterization of edges and thus estimation of importance are affected by a number of factors that can be related to the applied methods of data acquisition and of analysis~\cite{geier2017a,sanz2018}.
These confounding factors can lead to ambiguities, and we expect that our edge-centrality-based network decomposition can help to avoid misinterpretations that result from ambiguities. 

We here follow previous studies~\cite{schindler2008a,bialonski2011b,geier2015,geier2017a,stahn2017} and pursue a sliding-window approach which allows for a time-resolved analysis of characteristics of evolving weighted functional brain networks. 
To this end, we split the offline bandpass-filtered (\unit[1--45]{Hz}) EEG time series into consecutive non-overlapping windows of \unit[2.5]{s} duration each (corresponding to $T = 500$ data points) and normalize data from each sensor (vertex) and each window to zero mean and unit variance.
In order to derive edges, we estimate the strength of interaction between any pair of vertices $k$ and $l$ --~regardless of their anatomical connectivity~-- using an established data-driven method for studying time-variant changes in phase synchronization in EEG time series. 
The mean phase coherence~\cite{mormann2000} is defined as
\begin{equation}
 R_{kl} = \left| \frac{1}{T} \sum_{t=0}^{T} \exp \iu (\Phi_k(t) - \Phi_l(t))\right|,
\end{equation}
where $\Phi_k(t)$ are the instantaneous phases (here derived with the Hilbert transform) of the EEG time series from vertex $k$. 
$R_{kl}$ is confined to the interval $[0,1]$, and $R_{kl}=1$ indicates fully synchronized systems. 
In the resulting weighted snapshot network, self-loops are excluded.

\begin{figure*}[htp]
\includegraphics[width=1\textwidth]{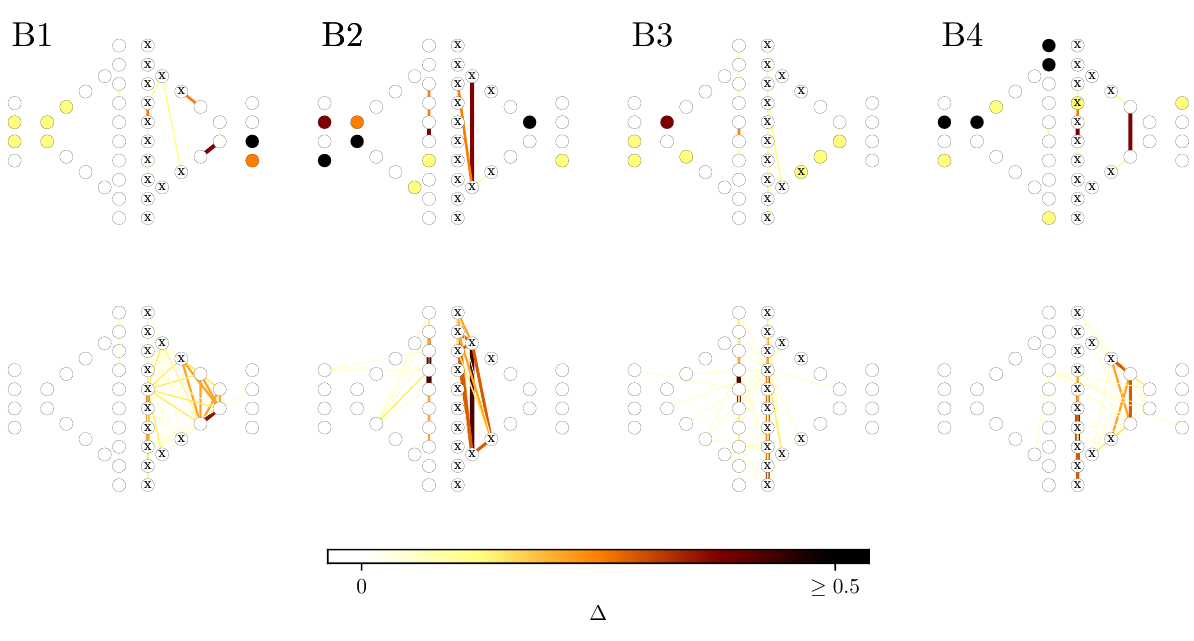}
\caption{Top: Most important edges and vertices (top) and most important web (rank 1, bottom) projected onto a schematics of implanted sensors shown in Fig.~\ref{fig:fig12}. 
The web was derived with the \CCE-based decomposition of the snapshot networks. 
Color (vertices and edges) and line thickness (edges) encode the relative amount of time $\Delta$ during each block for which a vertex/edge is rated as most important. 
Sensors (vertices) marked with `x' record from the clinically defined seizure onset zone.
}
\label{fig:fig13}
\end{figure*}

We begin by investigating the importance of edges and vertices in evolving functional brain networks 
using closeness centrality (we note that we achieved similar findings with other centralities). 
Since importance in such networks can vary strongly over time~\cite{geier2015,geier2017a,geier2017b}, we partition the recording into four blocks of equal duration, with each block containing the data from 13 consecutive snapshot networks, and report aggregated values of importance characteristics.
In the upper part of Fig.~\ref{fig:fig13}, we present the most important vertices and edges together with their temporal stability during each block's duration. 
We observe that most important vertices, and particularly those with a comparably high temporal stability, indicate brain regions outside the SOZ and even from the opposite brain hemisphere to be most relevant for seizure dynamics~\cite{geier2015}.  
In contrast, most important and temporally most stable edges do not connect most important vertices and point to interactions close to and within the SOZ as most relevant for seizure dynamics. 
Interestingly, they also point to homologous regions in the opposite brain hemisphere to be involved in the spreading of seizure activity.

In the lower part of Fig.~\ref{fig:fig13}, we present our findings obtained from applying our edge-centrality-based network  decomposition, which leads to \NWebs=650 webs, on average, with about 10 edges merged into the most important web $W_1$ (web with rank 1). 
The vast majority of temporally most stable edges point to interactions close to and within the SOZ as most relevant for seizure dynamics during most pre-seizure, seizure, and post-seizure phase.

Although these findings need to be validated on a larger database, they indicate that characterizing important edges or groups thereof in evolving functional brain networks can help to improve understanding of the complicated spatial-temporal dynamics of the epileptic process.

\section{Conclusions}
\label{sec:Conclusion}

We modified various, widely used centrality concepts for vertices to those for edges, in order to find which edges in a network are important between other pairs of vertices.
We also proposed an edge-centrality-based network decomposition technique to identify a bottom-up hierarchy of sets of edges or webs, where each web is associated with a different level of importance.
We investigated characteristics of the proposed edge centralities using paradigmatic binary and weighted network models.
We could demonstrate that despite the close relationship between importance of edges and vertices, edge centralities provide additional information about the network constituents for various topologies.

With the aim to identify important edges and important webs, we applied the novel concepts to a widely used benchmark model in social network analysis as well as to evolving epileptic brain networks.
Our edge-centrality-based network decomposition allowed us to identify a hierarchy of webs associated with levels of importance, which we consider advantageous particularly for those situations in which the most important edge can not be identified unambiguously~\cite{liao2017}.	
Our decomposition technique might also be of interest for identifying communities in networks~\cite{fortunato2016}. 
If these communities are considered a grouping of vertices, these vertices are allowed to be associated with multiple communities.
Our decomposition might also allow an alternative definition of communities, considering them as a grouping of edges.

In the majority of cases our decomposition technique identified a star-like structure as the most important web. 
Given that such structures are considered relevant for synchronization and percolation phenomena on complex networks~\cite{boccaletti2016b}, our edge-centrality-based network decomposition technique could help gaining deeper insights into such phenomena in real-world networks. 

Eventually, and with an eye on the inference of networks from multivariate time series, we are confident that the concepts and methods proposed here can help to improve characterization of networks through a data-driven identification of important edges or important webs.

\section*{Acknowledgements}
The authors would like to thank Thorsten Rings, Kirsten Simon, Sarmed Hussain, and Michael Roskosch for interesting discussions and for critical comments on earlier versions of the manuscript. 
This work was supported by the Deutsche Forschungsgemeinschaft (Grant No: LE 660/7-1).

\end{document}